 \documentclass[sigconf,screen]{acmart}

\AtBeginDocument{%
  \providecommand\BibTeX{{%
    \normalfont B\kern-0.5em{\scshape i\kern-0.25em b}\kern-0.8em\TeX}}}
\usepackage{microtype}
\usepackage{amsmath}
\usepackage{listings}
\usepackage{xcolor}
\usepackage{graphicx}
\usepackage{tabulary}
\usepackage{color}
\usepackage{colortbl}
\usepackage{multirow}
\usepackage{multicol}
\usepackage{booktabs}
\usepackage{float}
\usepackage{textcomp}
\usepackage{tcolorbox}
\usepackage{tikz,mathpazo}
\usepackage{makecell}
\usepackage{tabularx}
\usepackage{xspace}

\newcommand*{\circled}[1]{\lower.7ex\hbox{\tikz\draw (0pt, 0pt)%
        circle (.5em) node {\makebox[1em][c]{\small #1}};}}
\tikzset{
	state/.style={
		rectangle,
		rounded corners,
		draw=black, very thick,
		minimum height=2em,
		inner sep=2pt,
		text centered,
	},
}

\def\approach{{\sf FREPA}\xspace}
\def\tool{{\sf AeroReq}\xspace}
\def\FormalLanguageName{{\sf AASRDL}\xspace}
\def\githubTool{https://github.com/fengjincao/AeroReqDemoForFSE}

\begin{document}

%
\title[FREPA: Formal Requirement Engineering Platform in Aircraft]{FREPA: An Automated and  Formal Approach to Requirement Modeling and Analysis in Aircraft Control Domain}

\author{Jincao Feng  $^{1, \dagger}$, \ Weikai Miao$^{1, \dagger, *}$, \ Hanyue Zheng$^{1}$, \ Yihao Huang$^{1}$, \ Jianwen Li$^{1}$, \qquad\qquad\ Zheng Wang$^{3}$, \ Ting Su$^{1}$, \ Bin Gu$^{3,*}$, \ Geguang Pu$^{4,*}$, \ Mengfei Yang$^{5}$ , \ Jifeng He$^{1,2}$}
\thanks{
$^\dagger$ Both authors contributed equally to this research.\\
Jincao Feng's email: jincaofeng@foxmail.com \\
$^*$ Corresponding authors. E-mail:~{wkmiao@sei.ecnu.edu.cn, gubinbj@sina.com, ggpu@sei.ecnu.edu.cn}}
\affiliation{\institution{$^{1}$East China Normal University, China \ \ 
$^{2}$Shanghai Key Lab of Trustworthy Computing, China
 }
}
\affiliation{\institution{$^{3}$ Beijing Institute of Control Engineering, China\ \
$^{4}$Shanghai Trusted Industrial Control Platform Co., Ltd, China
}
}
\affiliation{\institution{$^{5}$China Academy of Space Technology, China\ \ 
}
}

\renewcommand{\shortauthors}{J.Feng, W.Miao, H.Zheng, Y.Huang, J.Li, Z.Wang, T.Su, B.Gu, G.Pu, M.Yang and J.He}
\renewcommand{\authors}{Jincao Feng, Weikai Miao, Hanyue Zheng, Yihao Huang, Jianwen Li, Zheng Wang, Ting Su, Bin Gu, Geguang Pu, Mengfei Yang, Jifeng He}

\begin{abstract}
Formal methods are promising for modeling and analyzing system requirements. However, applying formal methods to large-scale industrial projects is a remaining challenge. The industrial engineers are suffering from the lack of automated engineering methodologies to effectively conduct precise requirement models, and rigorously validate and verify~(V\&V) the generated models. To tackle this challenge, in this paper, we present a systematic engineering approach, named Formal Requirement Engineering Platform in Aircraft~(\approach), for formal requirement modeling and V\&V in the aerospace and aviation control domains. \approach is an outcome of the seamless collaboration between the academy and industry over the last eight years. The main contributions of this paper include 1) an automated and systematic engineering approach \approach to construct requirement models, validate and verify systems in the aerospace and aviation control domain, 2) a domain-specific modeling language \FormalLanguageName to describe the formal specification, and 3) a practical \approach-based tool \tool which has been used by our industry partners. 
We have successfully adopted \approach to seven real aerospace gesture control and two aviation engine control systems. The experimental results show that \approach and the corresponding tool \tool significantly facilitate formal modeling and V\&V in the industry. Moreover, we also discuss the experiences and lessons gained from using \approach in aerospace and aviation projects.
\end{abstract}


\maketitle
\section{Introduction}
Requirement modeling and analysis play a crucial role in developing high-quality software. 
A sufficient and precise requirement with rigorous validation and verification (V\&V) can significantly improve system design and implementation. Unfortunately, the mainstream approaches to requirement modeling remain the way to capture user intentions with natural language, which is widely used in today's software enterprises, from small companies to giant ones. Engineers usually have to spend a large amount of time inspecting the requirement documents to ensure their consistency and unambiguity, yet the effectiveness and efficiency of such manual work are far from satisfactory. The drawbacks of describing requirements with natural language include the lack of precision and the difficulty in automated V\&V. 

Both academics and industrial communities agree that rigorous system modeling and analysis can significantly contribute to the software quality~\cite{clarke1996formal,Kelly1997FormalNasaGuidebook, chechik2001automatic, newcombe2014use, bjorner2015checking,Jin2018Requirement, chudnov2018continuous,backes2019reachability}. Meanwhile, some industrial standards, e.g., the DO-333 for the aviation software~\cite{rtca2011333}, also require the application of formal methods through the development of safety-critical software. Modeling and analyzing system requirements with formal methods in software development life cycle are now considered as a promising solution to ensure software quality. 

Generally speaking, formal methods can extract a formal model from a specification, on which the rigorous mathematical proofs can be conducted for the V\&V purpose~\cite{jacky1997way,abrial2010modeling,abrial2005b}. Moreover, such a process can be fully automated. Unfortunately, applying formal methods to requirement modeling and V\&V in the industry remains a challenge, the main reasons for which are summarized as follows.

\begin{itemize}
    \item \textbf{Lacking dedicated notations for requirement modeling.}
    Most existed formal languages are too general to solve the domain-specific problems, and
    the dearth of the strong mathematical background makes understanding these formal notations difficult for most industrial engineers~\cite{clarke1996formal,miao2012formal}. 
    A proper formal language should not only precisely definite system functions and domain features, but also have these notations intuitive enough for users to easily understand them.
    
    \item\textbf{Lacking systematic approaches for requirement modeling and V\&V.} The truth is that even though industry practitioners want to apply formal methods, they do not know how to use formal methods in practice~\cite{wagner2019status}, i.e., how to build a formal requirement specification and what techniques can be adopted for requirements V\&V. In most cases, practitioners have already had documented requirements, under which situation it is difficult to write a formal specification from scratch. The engineers need a systematic approach to help them with writing formal requirement specifications to achieve formal V\&V without touching unnecessary complicated formal-proof tasks.  
    
    \item\textbf{Lacking  automation.} The lack of a powerful supporting tool is a severe problem for applying formal methods in industries. There are commercial tools that can provide formal verification, like SCADE~\cite{web2019Scade},  these tools are however more useful at the stage of system design rather than requirement modeling. For most industrial practitioners, mature tools for formal requirements modeling and V\&V are therefore highly in demand.
\end{itemize}

\begin{figure}[!t]
    \centering 
    \includegraphics[width=0.9\columnwidth]{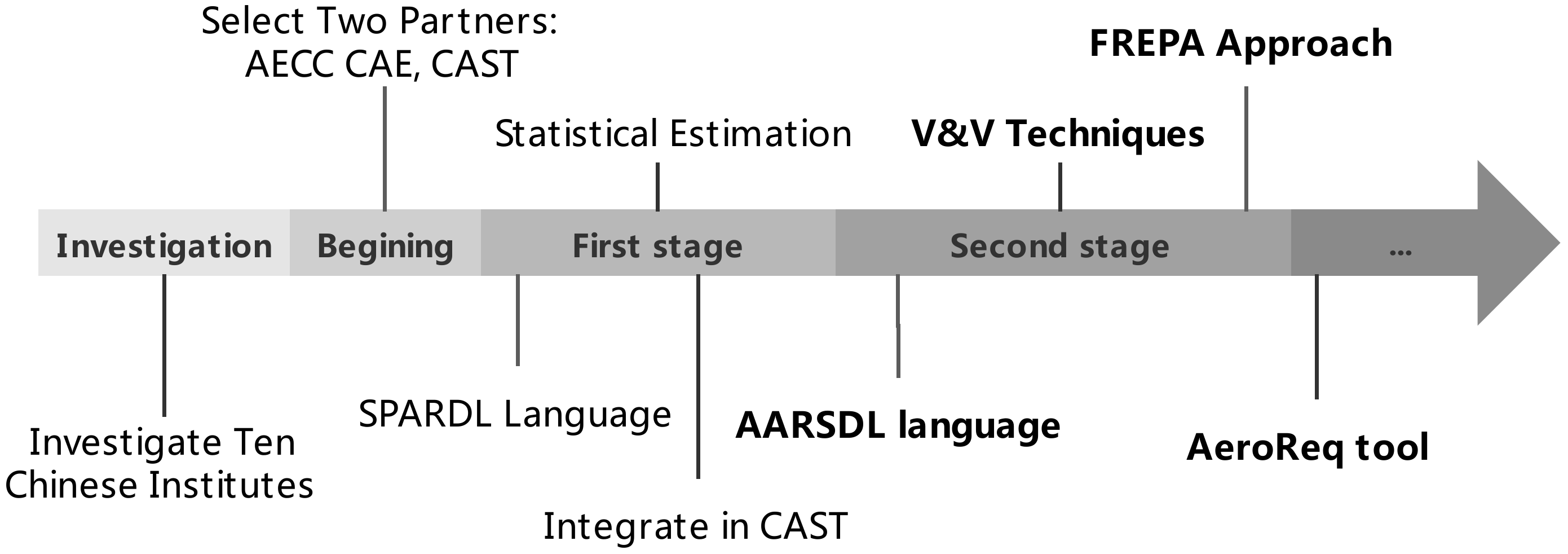}
    \caption{Timeline of the industry-collaborated projects}
    \Description{Project process from 2012 to now.}
    \label{fig:timeline}
\end{figure}

To tackle the above challenges, a promising solution is to develop a domain-specific engineering approach that integrates formal requirement modeling and V\&V techniques with tool support. Based on this motivation, we started a project from 2012 to conduct a comprehensive approach, which can support the formal and automated requirement modeling and analysis for the aerospace and aviation domains. As shown in Figure~\ref{fig:timeline}, we investigated more than 10 Chinese institutes in the domains. The software development in these institutes follows the trivial way to write requirements by simply using natural language. 
In the beginning, we selected two institutes as our research partners, one of which is Aero Engine Corporation of China Commercial Aircraft Engine Co. LTD~(AECC CAE), the largest aircraft engine producer of China. The other one is the China Academy of Space Technology~(CAST), the main spacecraft manufacturer in China. Two joint research groups were established respectively, in which members consist of researchers from East China Normal University and engineers from AECC CAE and CAST. 

At the first stage, according to an extensive study on the domain knowledge and philosophy of aircraft software development, we first presented a formal notation language, named SPARDL~\cite{wang2013MDM}, to capture the features of the aircraft control system. The formal notations defined in SPARDL are user-friendly and robust enough to represent the specific features in the aircraft control domain. Industrial practitioners can thus easily get used to modeling their requirements with SPARDL rather than other heavy-weight formal languages. After that, we developed a requirement analysis approach with V\&V techniques like statistical estimation~\cite{wang2012stochastic}. To apply the approach to industrial projects, we integrated the techniques into a prototype tool for the practitioners. Based on the engineers' feedback, we continuously improved the prototype until it can be successfully integrated into the software development process of CAST.

In the second stage of this project, before facilitating the application of the approach in real projects, we try to introduce the classic methodologies in academics like AADL~\cite{feiler2006sae} to the industry. However,  the practitioners think it is difficult to learn the formal notation and language, besides these methods don't fit their existing development process, they want to use an approach which can 
be easily integrated in. Thus, we generalized SPARDL and the relevant techniques for AECC CAE, achieving a more general and robust lightweight formal requirement modeling language Aerospace and Aviation Software Requirement Description Language~(\textbf{\FormalLanguageName}). Specifically, we develop a new systematic approach Formal Requirement Engineering Platform in Aircraft~(\textbf{\approach}) for \FormalLanguageName modeling and V\&V. 

\approach provides both the automated specification extraction and formal V\&V techniques including diagram-based review, simulation, statistical estimation, and test case generation under the MC/DC coverage criteria. We design the tool \tool to support the usage of \approach.~\approach was successfully applied to seven aerospace control systems from CAST, as well as two airplane engine control systems from AECC CAE. On average, tens of requirements errors including 5-10 fatal ones were detected for each system. The time cost for modeling and V\&V was reduced by at least 50\%, from 4-6 months to 2-3 months. The results show that \approach significantly improves the requirement modeling and analysis for our industrial partners. Our contributions are summarized as below:

\begin{figure*}[!h]
    \centering 
    \includegraphics[width=2\columnwidth]{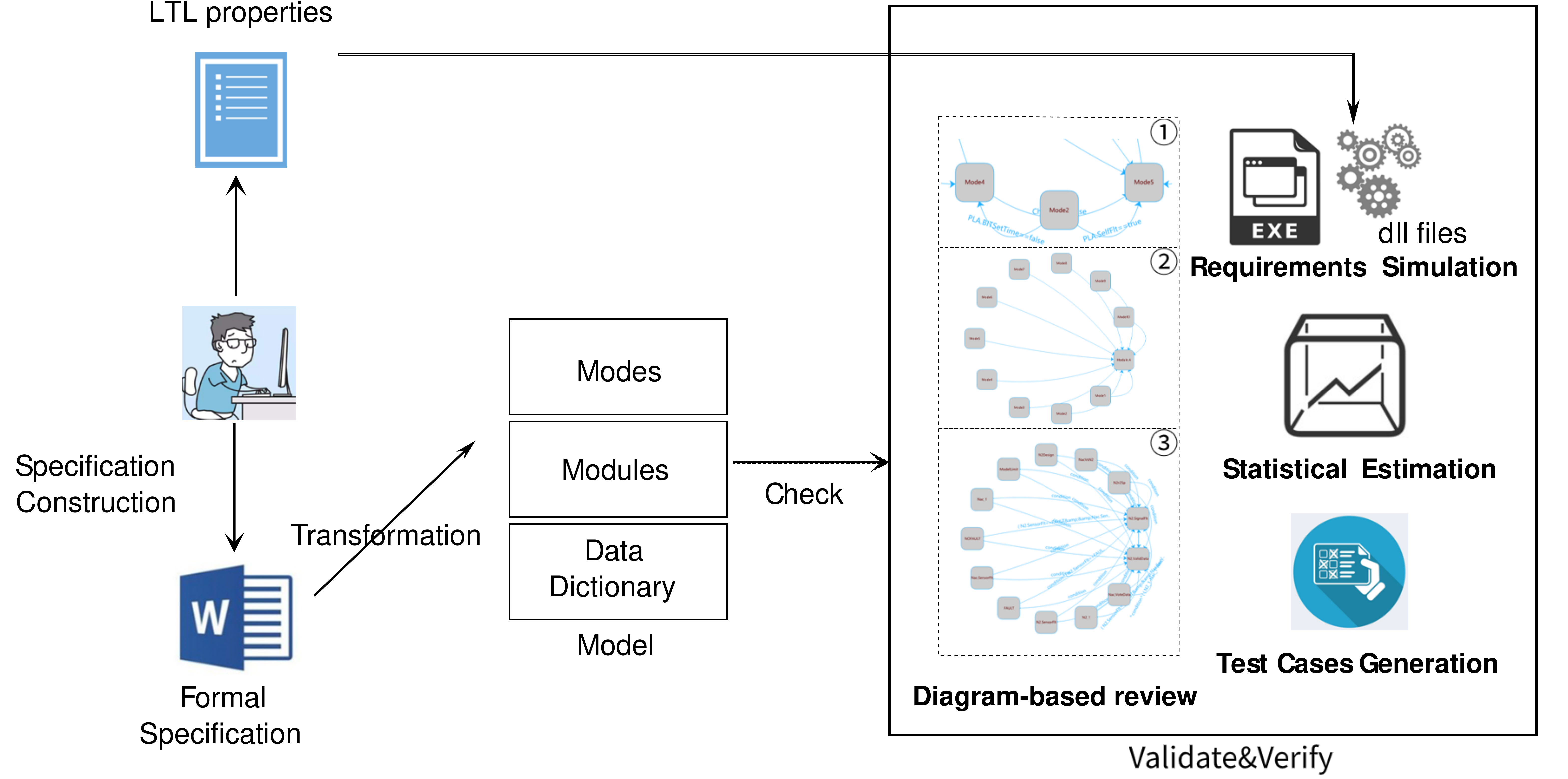}
    \caption{Framework of \approach}
    \label{fig:FEA_Framework}
      \vspace{-10pt}
\end{figure*}

\begin{itemize}
    \item A light-weight requirement modeling language \FormalLanguageName that supports aerospace and aviation domain features.  
    \item A systematic engineering approach \approach to the formal requirement modeling and V\&V for industrial practitioners in the aerospace and aviation domains;
    \item A tool \tool that supports the automation of \approach; 
\end{itemize}

In summary, we provide a systematic and automated solution to mitigate the challenges raised from the industry in terms of formal requirement modeling and V\&V. In our strategy,~\FormalLanguageName introduces a~\emph{domain-specific} modeling language with sufficient informal notations to construct the specification, \approach presents a systematic engineering framework to integrate modeling and V\&V technologies, and provides the automated analysis tool \tool for the requirements. Besides, we also share our experiences and lessons gained from using~\approach in real projects of the domains.

This paper is organized as follows.  Section~\ref{sec:framework} describes the main framework of~\approach. Sections~\ref{sec:formal_modeling} illustrates explicitly the techniques used in \approach.  Section~\ref{sec:exp} presents our experimental results. Section~\ref{sec:experience_lesson} summarizes the experience and lessons learned from the projects.Section~\ref{sec:rel_work} discusses related work. Finally, Section~\ref{sec:conclusion} concludes the paper.

\section{Framework of \approach}\label{sec:framework}
\approach consists of  formal modeling and V\&V techniques, whose framework is shown in Figure~\ref{fig:FEA_Framework}. To model the requirement, the engineers first manually document requirements and then construct their specification and properties with the help of the \FormalLanguageName template. After that, the models can be generated automatically, and then the V\&V technologies can be performed based on the generated model.

To perform the automated specification extraction from initial requirements, we provide a Microsoft Word template to facilitate the requirement writing. Such a template contains the components and sections w.r.t the requirement for the practitioner to fill in. To construct the formal specification, the practitioner simply needs to~\textit{fill in the blanks} of the template.

The specification is then transformed into an \FormalLanguageName model for V\&V, which consists of four major technologies:~\textit{diagram-based validation},~\textit{requirement simulation},~\textit{statistical estimation}, and~\textit{test case generation under the MC/DC coverage criteria}~\cite{yu2006mcdc}. Diagrams are generated from the formal model to visualize the system functionalities such that the practitioner can check whether the diagrams conform to their intentions. Requirement simulation checks some critical system behaviors by simulating formal model under certain inputs. To verify the safety-critical functional properties w.r.t the system,~\approach supports the statistical estimation. Rigorous requirement-based testing required by the related industrial standards (e.g., DO-333 for aviation software) are also supported.~\approach offers automated test-case generation under the MC/DC coverage criteria. 

\approach is completely implemented in our tool~\tool and we have already made our tool demo available on GitHub\footnote{\githubTool}. In the following, we briefly introduce the features of the formal modeling and V\&V techniques in \approach.

{\bf Modeling Requirements via a Word template.} Since most of our industrial partners are accustomed to building requirements with Microsoft Word, we provide a Word template such that they can write requirements under the guidance of keywords and blocks predefined in the template. Those predefined notations are represented in a structured way to precisely specify the software's domain features. Writing requirements with our provided template enables our partners to focus on the expected functions and simply fill the expressions into the template rather than dealing with complex mathematical notations. As soon as the template filling is completed, our tool \tool can automatically transform the textual requirements into a formal model for further V\&V procedures.

{\bf Formal Validation and Verification (V\&V).} Procedures are operated over the generated \FormalLanguageName model from the provided template. There are four major technologies integrated into \approach: the diagram-based review, requirement simulation, statistical estimation, and automated test-case generation under the MC/DC coverage criteria, which work together to achieve the goal of formal V\&V. 
\begin{enumerate}
  
   \item \emph{Diagram-based review.} We provide three diagrams w.r.t the constructed model for user review, i.e., the \textcircled{\oldstylenums{1}}\emph{mode transition}, \textcircled{\oldstylenums{2}}\emph{module relation}, and \textcircled{\oldstylenums{3}}\emph{variable dependency} diagrams. These diagrams visualize the relations among system modes, modules in each mode, and the variable interactions of each module, respectively. The practitioners thus can validate whether the requirements conform to their informal specifications in vision.
    \item \emph{Requirements simulation.} In \tool, the formal model can be dynamically executed for the simulation purpose. Comparing the simulated and expected results given by the domain experts, we can check whether the system's behaviors are correctly captured in the requirement.
    
    \item \emph{Statistical estimation.} We utilize this technique to verify whether the requirement model satisfies the properties under a given probability threshold.
    
    \item \emph{Automated test-case generation.} Since the requirements are formally defined, the automated and requirement-based test-case generation under the MC/DC coverage criteria becomes possible. Although testing is not necessarily an activity of requirement V\&V, it is required by some industrial standards such as the DO-333 for the aviation systems. \tool is designated to make use of the verified formal model for future testing purposes.

\end{enumerate}

\section{APPROACH}\label{sec:formal_modeling}
\subsection{Template-guided Formal Specification Construction} 
From our investigation, common features exist in both aerospace and aviation control domains. These common features inspire us to design a general modeling language that can describe the system requirements in both domains.
In summary, three major features in both the aerospace and aviation control systems are listed as follows.

\begin{itemize}
    \item \textbf{Mode-based:}  The control software is a~\textit{mode transition} system, which always runs in some mode to perform corresponding functionalities. The system can transit to another mode if the transition condition is triggered.
    \item \textbf{Computation-oriented:}  The control software usually focuses on decision and mathematics computations.
    \item \textbf{Period-driven:} The control software performs functionalities under the restriction of given periods. For example, in the \emph{cruise} mode, a satellite has to check the value of the outside temperature every 5 seconds.
\end{itemize} 

\begin{figure}[!t]
    \centering 
    \includegraphics[width=\columnwidth]{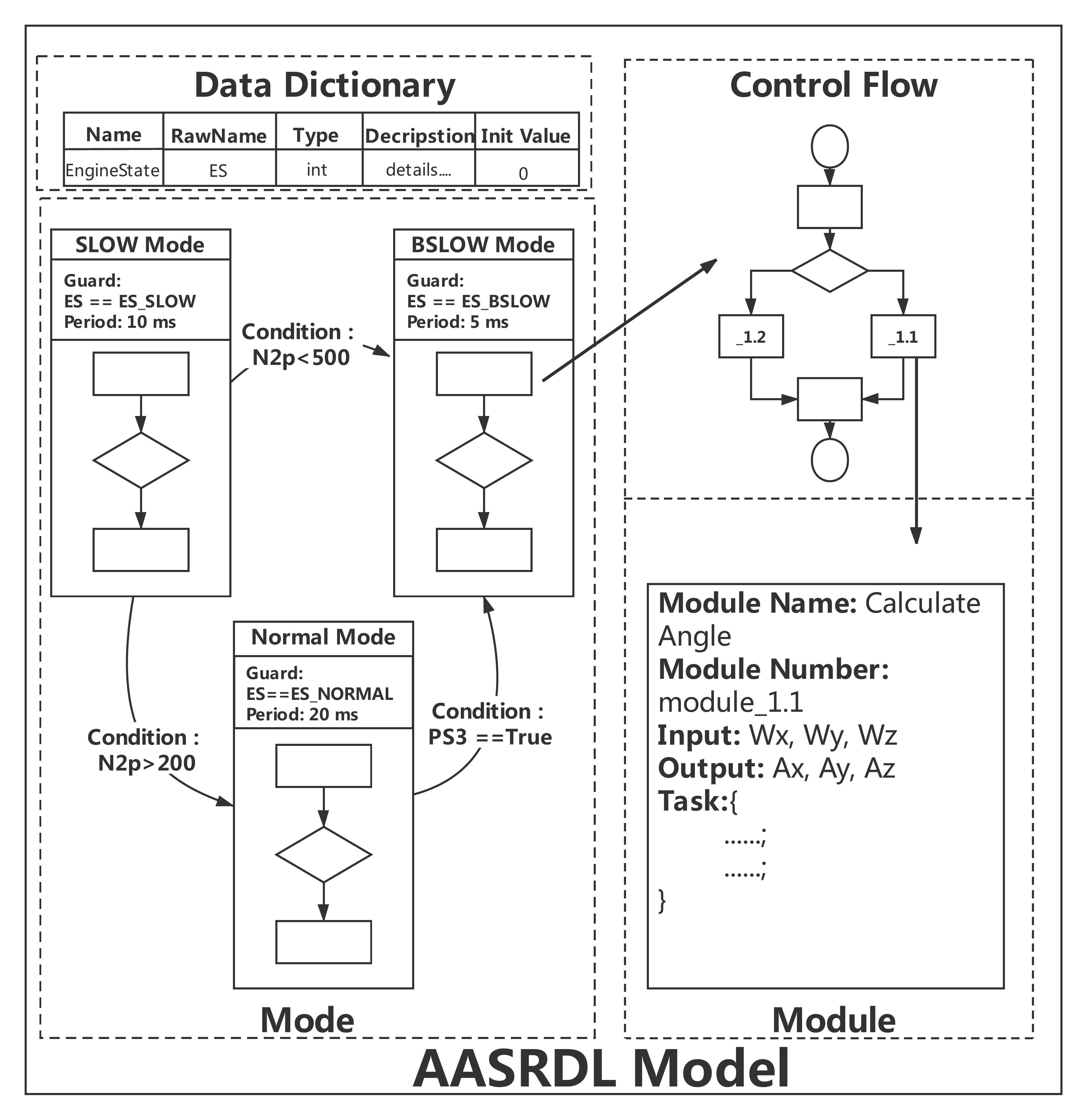}
    \caption{Structure of the \FormalLanguageName Model}
    \vspace{-15pt}
    \label{fig:AASDL_model}
\end{figure}

We thus present Aerospace and Aviation Software Requirement Description Language~(\FormalLanguageName), a generalization of SPARDL~\cite{wang2013MDM}, which is robust to model systems in both domains.
The schema of the language is illustrated as follows.
\begin{align*}
    & Model ::= (\{Mode\}, \{Module\}, DataDict) \\
    & Mode ::= (Name, Guard, \{Procedure\}, \{Transition\})\\
    & Procedure ::= (Period, Control Flow) \\
    & Transition=(Priority, TargetMode, Condition, Action) \\
    & Module ::= (Name, V_{in} \subseteq DataDict, V_{out}\subseteq DataDict, Task)\\
    & DataDict ::= \{V|V=(Name, Type, InitValue, Min, Max)\}
\end{align*}
In the above, \textit{Guard} and \textit{Condition} are first-order predicates whose evaluations can be \textit{true} or \textit{false}. We assume readers are familiar with such a concept and ignore the details here. 
The structure of an \FormalLanguageName model is shown in Figure~\ref{fig:AASDL_model}.
An \FormalLanguageName model consists of a set of system \emph{mode}s and \emph{module}s as well as a data dictionary \emph{DataDict}. 
The system always runs in some \emph{mode}, in which the contained procedures are executed in order. 
Each \emph{procedure} of a \emph{mode} includes a \emph{period} and a \emph{Control Flow}, which is a set of C-style statements 
without the loop. The task of the \emph{procedure} is to execute the corresponding \emph{Control Flow} within the given period.
The mode \emph{transition} occurs when the \emph{condition} is evaluated to be \textit{true}, in which some \emph{action}~(same as \emph{Control Flow}), is executed for certain purposes. 
The \emph{mode} transits to the \emph{target mode} once the corresponding \emph{guard} is evaluated to be \textit{true}; otherwise, it has to rollback its \emph{action}.
A \emph{module} performs the functionalities in its \emph{task}~(same as \emph{Control Flow}). All \emph{module}s have to declare their input $V_{in}$ and output variables $V_{out}$, which are all precisely defined in \emph{DataDict}. A \emph{mode} calls its corresponding \emph{module}s to achieve the duty. 
The \emph{DataDict} defines all the variables that are used in the model. The information on the type, initial, minimal and maximal values for each variable are collected in \emph{DataDict}. 

The system maintains a global clock by default (not defined in the language syntax) to coordinate the running time of different procedures. Since this paper fills in the scope of the industrial track, we omit the language semantics, and the reader is referred to \cite{wang2013MDM} for more details.     

The practitioners usually suffer from building a formal specification from scratch due to the lack of guidelines. To tackle this problem,~\approach provides a template to guide the formal requirement modeling. In the template, the practitioner only needs to fill in the designated sections using~\FormalLanguageName expressions and predicates. The sections are specified by keywords such as~\textit{Mode Name}, which guides the practitioner to write the name of a system mode. 

\begin{figure}[!t]
    \centering 
    \includegraphics[width=1.0\columnwidth]{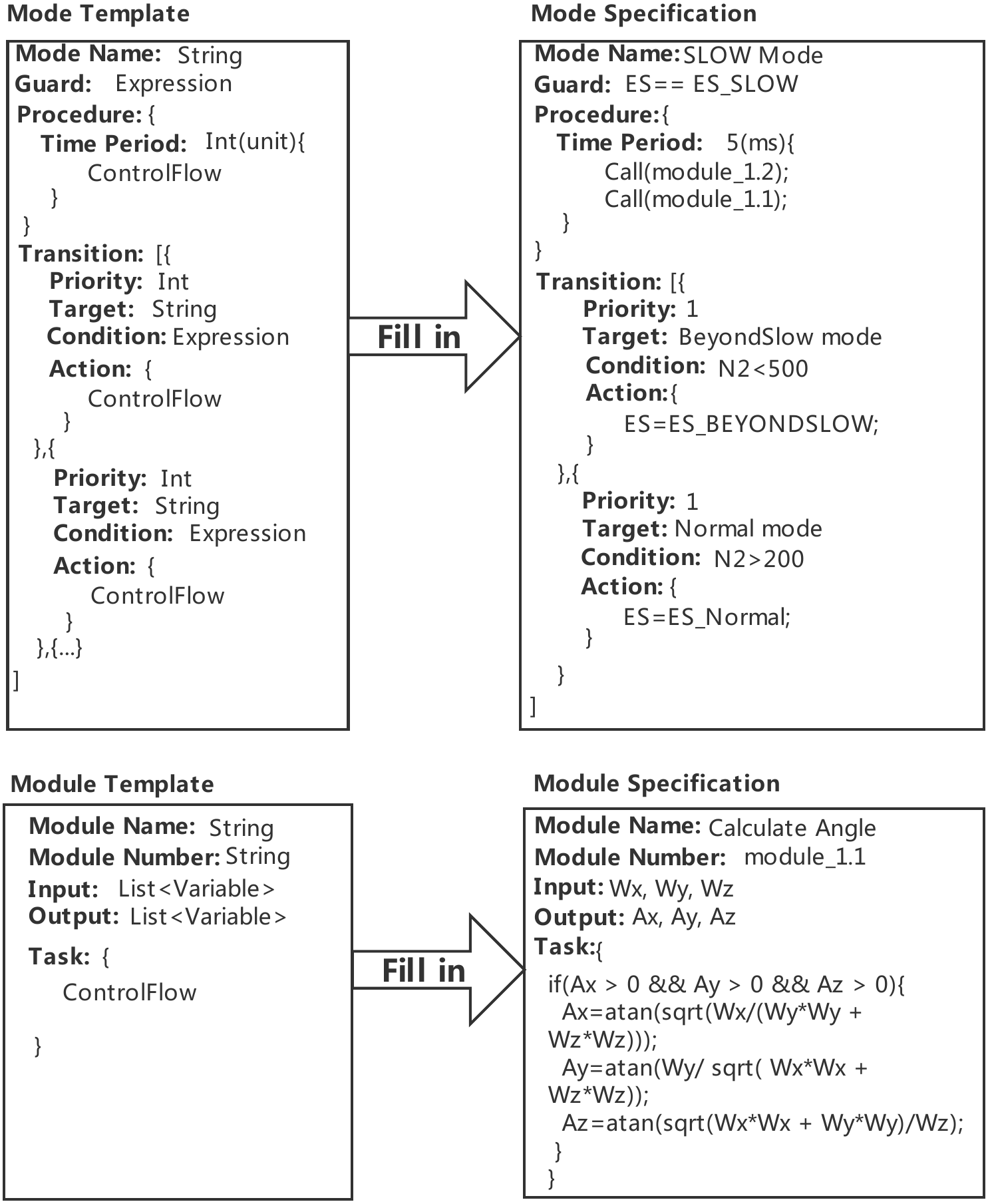}
    \caption{An example of the fill-in template process}
     \vspace{-15pt}
    \label{fig:formal_specification}
\end{figure}
Figure~\ref{fig:formal_specification} is an example to show how to fill the requirement in the template. The left of Figure~\ref{fig:formal_specification} is the template we provided, the right side of the figure is the specification. The keywords such as \textit{\sl Mode Name}, \textit{\sl Procedure} and \textit{\sl Transitions} explicitly define the components to describe the requirements.  In the specification, the keywords such as the "\textit{\sl Mode}", "\textit{\sl Procedure}", and the "\textit{\sl Transitions}" explicitly defines the components for describing the requirements. 

The practitioners can obtain the formal specification easily as soon as the filling is completed. As shown in the right of  Figure~\ref{fig:formal_specification} is the specification including one Mode and one Module. In this specification, the period of the \textit{SLOW Mode} is 5 milliseconds, which means that the computation functions including the~\textit{\sl moudule\_1.1, \sl module\_1.2} are executed every 5 milliseconds. The function \textit{\sl module\_1.1} performs its task and returns the value of its output variables~\textit{\sl Ax, Ay, Az}. If the condition \textit{\sl N2<500 } is true, the system transits to the \textit{\sl BEYONDSLOW}~\textit{Mode}. If the condition~\textit{\sl N2>200 } is true, the system transits to the~\textit{\sl NORMAL} \textit{Mode}. we will use this specification as our running example to motivate our approach. 

Once we get the specification, our tool \tool takes \emph{Antrl}~\cite{parr1995antlr} as the compiler for specifications. In that way, we can transform the specification into an \FormalLanguageName model  for the next formal V\&V technologies.

\subsection{Diagram-based Requirements Review}
\begin{figure}[!t]
    \centering 
    \includegraphics[width=1.0\columnwidth]{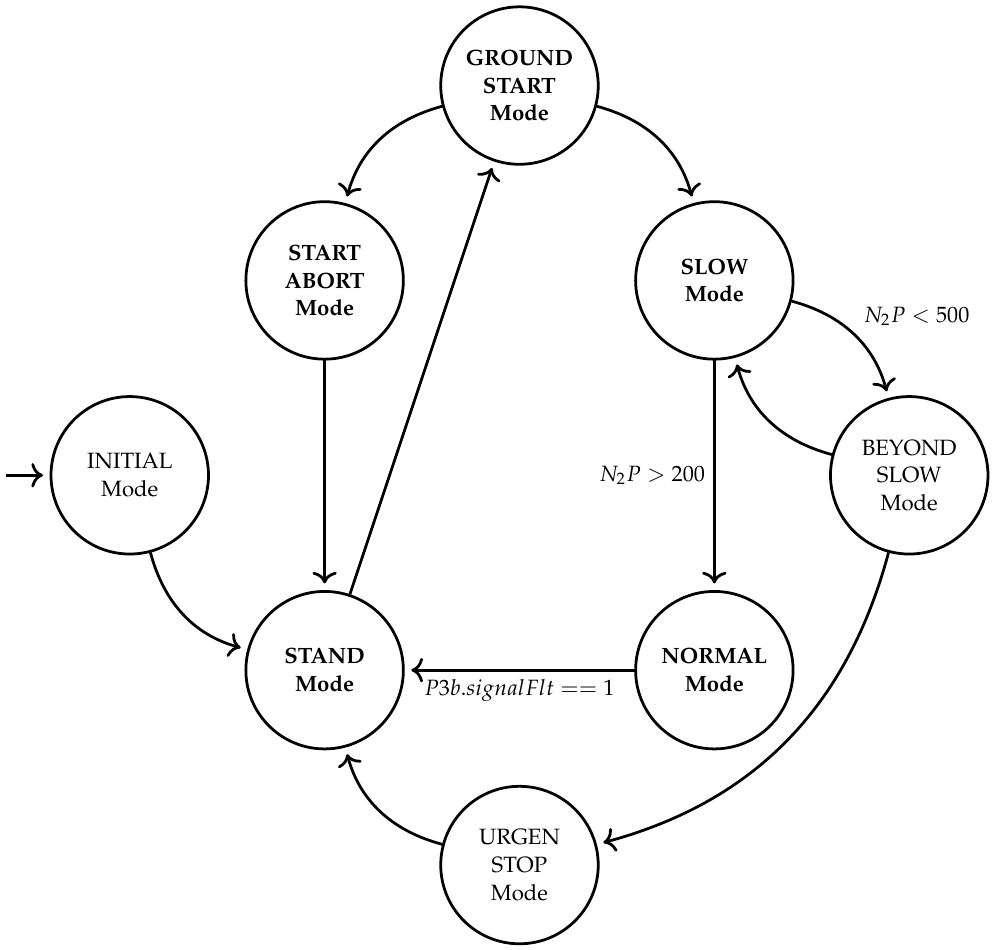}
    \caption{Mode Transition Diagram of the system}
     \vspace{-10pt}
    \label{fig:mode_diagram_en}
\end{figure}
Since industrial software requirements are usually written in a large number of documents, manually validating the textual specification is error-prone. To tackle this problem, \approach offers a diagram-based review method. The basic idea is generating diagrams based on the formal model to visualize the requirements. Then the practitioner reviews the diagrams to check whether the diagrams conform to their intentions. Since diagrams can visualize the overall requirements intuitively, such as reviewing on Graphic User Interface~(GUI) is more effective than that directly on the textual documents by the practitioners' experience. With a thorough review, some errors in the requirements can be found by the practitioner according to the domain knowledge.

In~\approach, we generate three diagrams from the model: the mode transition, module relation,  and variable dependency diagrams. Due to the limited space, we only focus on the explanation of how the mode transition diagrams work. Figure~\ref{fig:mode_diagram_en} describes a mode diagram that is generated from the specification of an airplane engine control system.
If the condition \textit{P3b.SignalFlt == 1} is true, the system transits from \textit{NORMAL Mode} to \textit{STAND Mode}. 
The requirement review for mode diagrams is performed according to the following criteria.

\begin{itemize}
    \item \label{criterion1} \textbf{Criterion 1: } Each transition in a mode should conform to the intended perceptions;
    \item  \label{criterion2} \textbf{Criterion 2: } Mode transitions should be exclusive;
    \item  \label{criterion3}\textbf{Criterion 3: } Every mode should be reachable.
\end{itemize}

Criterion 1 is a basic guideline for diagram-based review. The practitioners check whether the mode transitions conform to their expectations based on domain knowledge. For instance, the practitioners find that the system should not directly transit from the \textit{URGENSTOP Mode} to the \textit{NORMAL Mode}. 
Criterion 2 requires that a mode never satisfies more than one transition condition simultaneously.
Criterion 3 ensures that the system can transit from one mode to any other mode eventually. Checking criteria 2 and 3 are actually constraint solving problems. For instance, when checking whether two transitions are exclusive, search whether there exists at least one solution that makes both transition conditions true.

Here we run an example of exclusiveness checking. Let's consider two transition conditions of the \textit{SLOW Mode}. If the condition \textit{\sl N2<500} is true, the \textit{SLOW Mode} should transit to the {BEYONDSLOW Mode}. If the condition \textit{\sl N2>200} is true, the \textit{SLOW Mode} should transit to \textit{Normal Mode}. Obviously, \textit{\sl N2<500} and \textit{\sl N2>200} can both be true when \textit{\sl N2>200 $\&\&$ N2<500} is true. That means the two mode transitions are not exclusive. 

Our tool \tool integrates with Z3~\cite{de2008z3} as the constraint solver to check the exclusiveness of mode transitions. Analogously, we can check the reachability among different modes.

\subsection{Requirement Simulation}
Simulation is often used for the V\&V of system designs~\cite{de2012reporting} in commercial tools such as \textit{SCADE}. However, few simulation techniques exist in the requirement level because in many cases, requirements are not precise for automated simulation. To mitigate this issue, \approach formally defines the requirements such that automated simulation becomes possible. 
\begin{figure}[!t]
    \centering 
    \includegraphics[width=1.0\columnwidth]{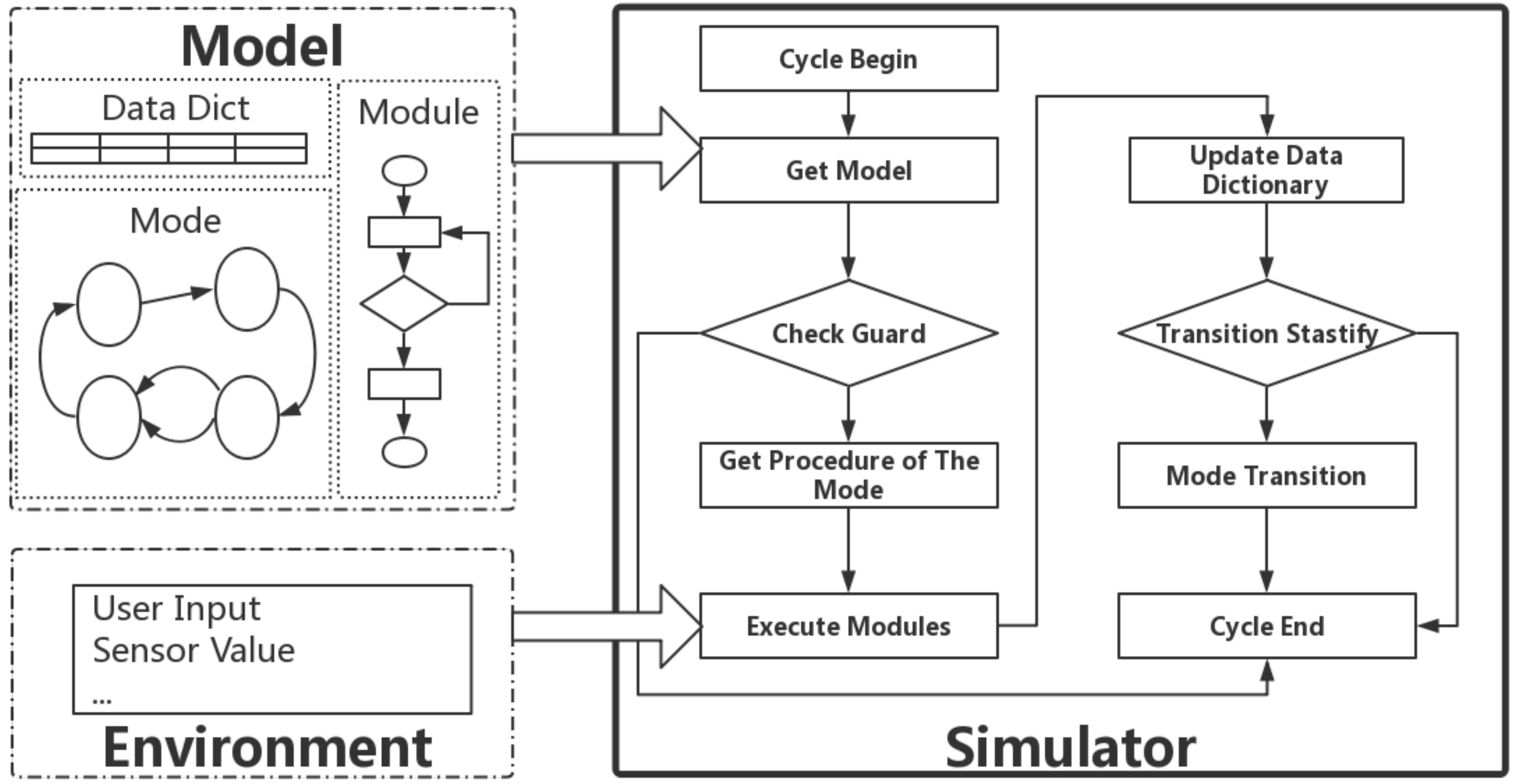}
    \caption{Overview of simulation.}
     \vspace{-10pt}
    \label{fig:Simulation}
\end{figure}

Indeed, we developed a corresponding simulator to simulate directly on the \FormalLanguageName models. Figure~\ref{fig:Simulation} shows the process of the simulation. The requirement simulation needs some external settings and input values provided by the domain experts Based on the model and environment, we can simulate our model periodically and review the results to verify whether the requirement is what we need.

The simulator starts from the initial mode and ends when the stop signal is received. In every cycle, our simulator first checks whether the \emph{Guard} of each mode is satisfied. If it is the case, the simulator simulates the procedure of the mode. Otherwise, the simulator will stop because of requirements errors. When the simulation of a mode is completed, we will check whether the \emph{condition} of the \emph{transition} is satisfied. The mode transition will occur at the end of the cycle.

We use the specification in Figure~\ref{fig:formal_specification} as an example. The simulator starts from the \emph{SLOW Mode}. If its guard condition \emph{Guard}($ES==ES\_SLOW$) is True, the modules \emph{module\_1.2}, \emph{module\_1.1} are called in the order in one period. When simulating \emph{module\_1.1}, we calculate the outputs' values~(e.g., Ax, Ay, Az) by using the input values and the control flow. When the procedure is finished, all the \emph{conditions} of the transition are not satisfied, the simulator will enter the next cycle and repeat its procedure. During the simulation, all the values of the output variables in every period are recorded in a file. Therefore, we can generate various charts for the result analysis.  

\subsection{Statistical Estimation}\label{sec:formal:se}
Model checking~\cite{Clarke2018model} is a classical technique to verify whether a model satisfies the given property. One popular language for specifying such properties is LTL~(Linear Temporal Logic)~\cite{Pnueli77}, which reasons over linear executions of system. The control software systems of aircraft and spacecraft are typical linear execution. Traditional model checking techniques, however, cannot be directly adopted, since the aircraft execution is also affected by the random environmental factors. Therefore, we utilize the \emph{statistical estimation} technique introduced in~\cite{wang2012stochastic} to check whether the model complies with the desired properties specified in LTL under a specific threshold. The \emph{statistical estimation} technique is a light-weight version of model checking~\cite{legay2010statistical}, that uses the statistical methodology to estimate the probability of a model satisfying the given property.

Given a model \textit{M} and a property \textit{p}, there are two major objectives for statistic estimation. The first objective is to estimate the probability of whether \textit{p} is satisfied on every execution path \textit{t} of \textit{M}. The other one is to judge whether the probability is bigger than a given threshold. The idea of statistical estimation is straightforward: we calculate the statistics of the successful and failed model executions during 
the simulation and finally return the probabilistic verification result. For instance, if a property can hold 99 times under 100 times of execution, the probability of this property to be held is 99\%.  

For example, a spacecraft will reach a stable state, \textit{i.e.},  the angle and the angular velocity must be smaller than a threshold after certain periods. We use \textit{\sl Ax, Ay, Az} to denote the angle and \textit{\sl Wx, Wy, Wz} for the angular velocity. The property that \textit{spacecraft will reach a stable state and then stay stable} can be described in LTL  as below: 
\begin{equation*}
    FG(\sqrt{A^{2}_{x} + A^{2}_{y} + A^{2}_{z}} \leqslant 0.1 \wedge \sqrt{W^{2}_{x} + W^{2}_{y} + W^{2}_{z}} \leqslant 0.01)
\end{equation*}

\subsection{Automated Test-Case Generation under the MC/DC Coverage Criteria}

\begin{table}[t]
    \small
    \centering 
    \caption{Test Cases Generated under MC/DC Coverage.}
    {
        \begin{tabular}{|c|c|}
            \hline
            Constraints	&Test Cases\\ \hline
            {Ax>0$\wedge$Ay>0$\wedge$Az>0  }&{Ax = 1; Ay = 1; Az = 1;}\\ \hline
            {!( Ax>0$\wedge$Ay>0$\wedge$Az>0)}&{Ax = 0; Ay = 1; Az = 1;}\\ \hline
            {!( Ax>0$\wedge$Ay>0$\wedge$Az>0)}&{Ax = 1; Ay = 0; Az = 1; }\\ \hline
            {!( Ax>0$\wedge$Ay>0$\wedge$Az>0) }&{Ax = 1; Ay = 1; Az = 0; }\\ \hline 
            
        \end{tabular}
        \vspace{-10pt}
    }
    
    \label{table:MCDC_table}
\end{table}
The DO-333 standard for aviation software claims that requirement-based coverage analysis has to be achieved with using formal methods. The purpose of requirements-based coverage analysis is to determine how well the implementation of the software requirements has been verified. Thus, the practitioners need the requirement-based test cases to achieve the complete coverage of each requirement, and to detect of unintended data flow relationship between the inputs and outputs as well as the dead code and deactivated code. To achieve this goal, We use dynamic symbolic execution~\cite{zhang2005constraint} implementation supports MC/DC coverage~\cite{zhang2014MCDC,zhang2018smartunit} from the specification. 

The idea to generate test cases is to collect the constraints and conditions of the modules control flow structures, and then compute the results that satisfy the constraints and conditions. These results are the pursued test cases.Given a specific condition statement \textit{S}, the conditions along the control flow path from the root node to \textit{S} are called \textit{constraints}. Then the test case generation is an SMT solving problem. Our tool \tool integrates Z3~\cite{de2008z3} as the solving engine. Taking \emph{module\_1.1} in Figure~\ref{fig:formal_specification} as the example, the \textit{S} in this module is that \textit{\sl Ax>0 $\&\&$ Ay>0 $\&\&$ Az>0}, the generated \textit{constraints}, and the generated test cases are shown in Table~\ref{table:MCDC_table}. 

\section{Experiments}\label{sec:exp}
\subsection{Experimental Setup}
\label{sec:eval_setup}

\noindent\textbf{Goals.} In this section, we introduce the performance of \approach on the two benchmark systems. We aim to show the effectiveness of \approach in the following subsections.

\noindent\textbf{Platform.}
Our tool \tool was running on a computer of CAEC, which has two processors~(2.70 GHz Intel(R) Core(TM) i5-6400 CPU) with 8 GB memory. The operating system on the computer is 64-bit Windows 7. For the projects of the aerospace control system, \tool  was running on a server in CAST, which has eight processors~(2.33 GHz Intel(R) Xeon(R) E5345 CPU) with 16 GB RAM. The operating system on the server is 32-bit Windows Server 2003.

\noindent\textbf{Industrial Benchmarks.}
As mentioned before, \approach has been applied to seven aerospace gesture control and two airplane engine control systems for formal V\&V. 
Based on the contract with our industrial collaborators, we only have the chance to use \approach from scratch on one of the systems each as the demonstration, and other systems that applied \approach to are completed by our industrial partners due to the privacy regulation.  
We thus introduce the evaluation results on these two demonstrative systems. The formal specification of the aviation system is a 128-page Word document with 10 system modes, 139 modules, and 1200 variables, while the specification of the aerospace system is a 145-page Word document with 14 system modes, 42 modules, and 420 variables.

\subsection{Results and Analysis}
\label{sec:eval_result}

\begin{tcolorbox}
\noindent\textbf{\emph{G1}}: How useful for the diagram-based review to improve the requirement analysis?
\end{tcolorbox}
We generated a 10-mode transition diagram, 152 module-relation diagrams,and 152 variable-dependency diagrams for the aviation system. A total of 99 errors were detected during that process, which includes before-use defined or uninitialized variables as well as the incomplete inputs and outputs of modules. 
In practical, the errors may be caused by some trivial mistakes, such as inconsistent capitalization or typos. By reviewing those generated diagrams manually, it helps to detect 9 errors, such as 4 wrong variable dependencies, \emph{circular dependencies} among variables, in which deadlocks may occur, and 2 nonexclusive and 3 unreachable mode transitions. Similarly, for the aerospace system, we generated a 14-mode transition diagram, 42 module-relation diagrams, and 42 variable-dependency diagrams. 17 errors were thus detected.
In general, generating diagrams from the specification helps to detect errors caused by imprecise requirements, while manually reviewing the diagrams helps to find logic errors. 

\begin{tcolorbox}\noindent\textbf{\emph{G2}}: How useful for the requirement simulation to detect requirement errors?\end{tcolorbox}
To evaluate the requirement simulation results, we compared the values of key variables computed by the simulator against the expected ones from simulink provided by our industrial partners. The motivation came from that any inconsistency between the simulated and expected values indicated the possibility of error occurrence. We selected the engine-starting process of the aviation system as an example to show the simulation results. The engine-starting process requires that the system should start from the \emph{initial} mode, then transit to the \emph{STAND Mode}, \emph{GROUND-START Mode}, \emph{SLOW Mode} in order. The relevant key variable is \textit{N2}, which represents the high-compressor-rotor speed of the engine. Our purpose was to simulate the whole process of the engine-starting to check whether \textit{N2} meets expectations. 

\begin{figure}[!t]
    \centering 
    \includegraphics[width=1.0\columnwidth]{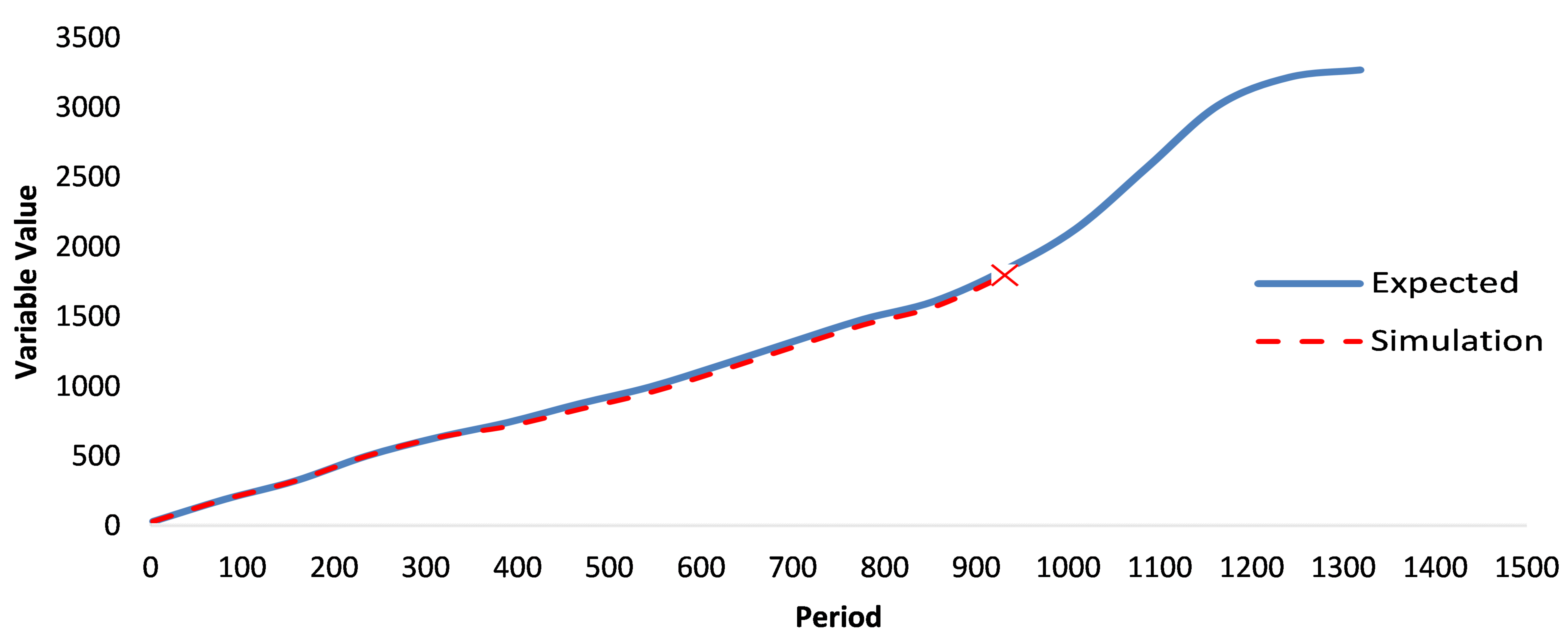}
    \caption{Unexpected simulation for the aviation system.}
    \label{fig:SimulationResult_Error}
\end{figure}

We detected 2 errors based on the simulation. Figure~\ref{fig:SimulationResult_Error} is the simulation result of the engine-starting process. The dotted curve has a breakpoint  approximately 1000 periods, which shows the simulation terminates with exceptions. 
By investigating log files, we found the error was caused by the abuse of variable \textit{LowOil} representing the low oil pressure value. The \emph{Guard} in the \emph{GROUND-START Mode} restricts that \textit{LowOil} cannot always below the threshold during a certain time. In the simulation, \textit{LowOil} does not satisfy the \emph{Guard} such that the transition to the \emph{GROUND-START Mode} cannot succeed. As a result, the simulation stops and throws an exception. 

\begin{figure}[!t]
    \centering 
    \includegraphics[width=1.0\columnwidth]{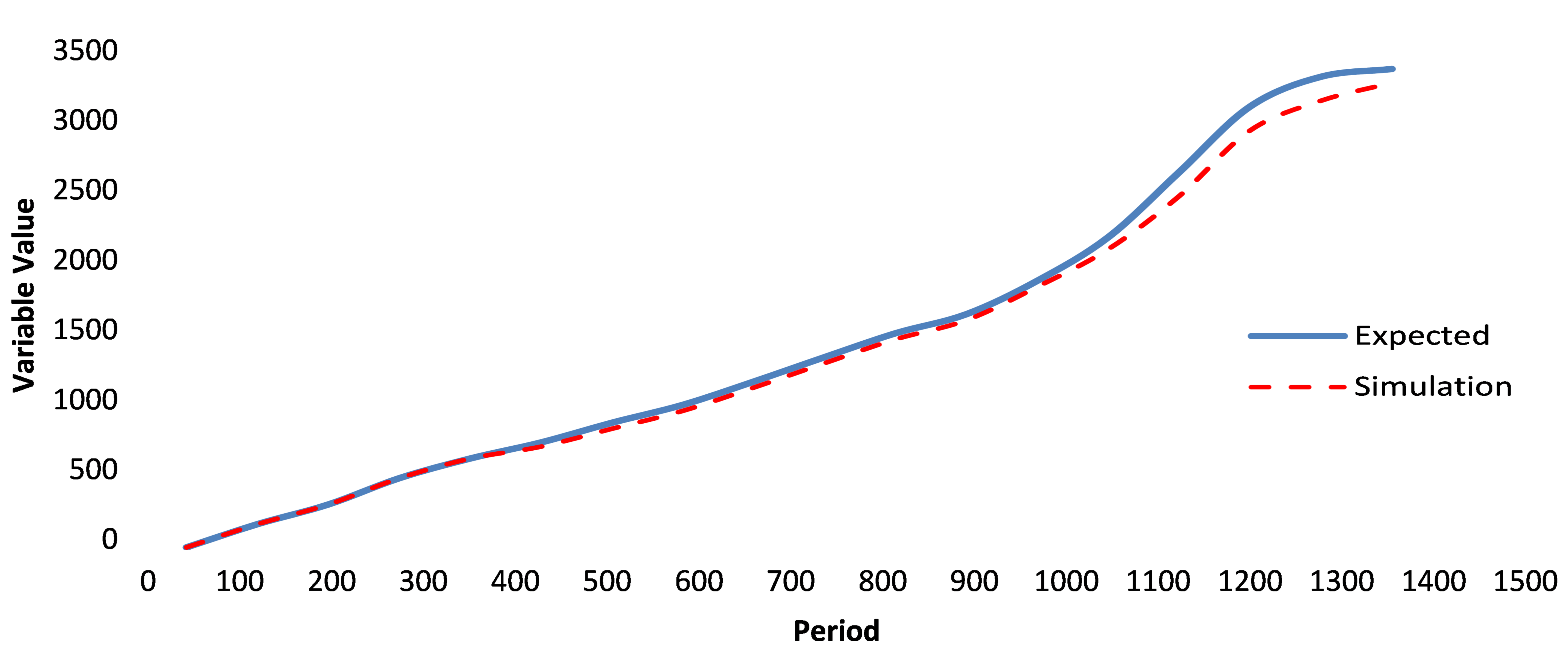}
    \caption{Normal simulation for the aviation system.}
     \vspace{-10pt}
    \label{fig:SimulationResult_Right}
\end{figure}

We re-simulated the engine-start process after fixing the problem discovered above. Figure~\ref{fig:SimulationResult_Right} shows the corresponding result, in which the simulation result is not strictly consistent with the expected one after 1000 periods. The practitioners hence double-checked the requirement and found that \textit{N2} is a \emph{float} type variable instead of the \emph{double} type originally given by the company. The small deviation at the beginning resulted in a much bigger difference after thousands of periods. 

In general, our simulation technique is helpful to find two categories of errors that frequently occur. The first one is the logical error, e.g., the incorrect computation. The second one is  inadequate requirements. For instance, the aerospace control system has a function for monitoring outside temperature. In the requirement, the practitioner did not define a \textit{reset} operation in this function. Therefore, the simulation stopped when the temperature reached the upper bound. 

\begin{tcolorbox}\noindent\textbf{\emph{G3}}: How useful for the statistical estimation to identify more requirement errors than human work?\end{tcolorbox}
We selected three properties that the engine control system should satisfy for the experiment. The corresponding explanations of the three properties are listed below. 
\begin{enumerate}
    \item\label{sec:result:rq3:p1} The system will eventually reach a \emph{stable} state and stays there forever.
    \item\label{sec:result:rq3:p2} The system starts from \textit{mode m2}, and then it will finally transit to \textit{mode m5} or \textit{mode m6} or \textit{mode m8}, and stay forever in one of these three modes.
    \item\label{sec:result:rq3:p3} When the system leaves \textit{mode m4}, it has to transit to \textit{mode m0}, and then to \textit{mode m1}, and finally to \textit{mode m2}.
\end{enumerate}

We then formulated these properties in LTL and utilized the statistical estimation introduced in~\cite{wang2012stochastic} to check whether the model complies with the desired properties. 
The property translation from natural language to LTL is conducted manually. We take property~\ref{sec:result:rq3:p1} as an example. The term ``stable'' means that the angle and the angular velocity must be smaller than the given thresholds. We can use mathematical formulas $\sqrt{A^{2}_{x} + A^{2}_{y} + A^{2}_{z}} \leqslant 0.1 \wedge \sqrt{W^{2}_{x} + W^{2}_{y} + W^{2}_{z} }\leqslant 0.01$ to express this property. And the term ``forever'' can be formally expressed by the temporal LTL operators, e.g., $X$, $F$, and $G$.
The following LTL formulas are the formal properties that are translated from the above natural-language ones.
\begin{small}
    \begin{align}
        & FG(\sqrt{A^{2}_{x} + A^{2}_{y} + A^{2}_{z}} \leqslant 0.1 \wedge \sqrt{W^{2}_{x} + W^{2}_{y} + W^{2}_{z}} \leqslant 0.01)\\
        & (ES=2) \wedge      FG(ES=5 \vee ES=6 \vee ES=8) \\
        &  G((ES=4) \Rightarrow (X (ES=0) \Rightarrow (X (ES = 1) \Rightarrow X (ES=2))))
    \end{align}
\end{small}

Suggested by the engineer, we set the simulation period from 0 to 5000. The \emph{semi-confidence interval} was set to 1\% ($\delta$ = 1\%). The \emph{confidence} was set to 5\% ($\sigma$ = 5\%). The time limit for property verification was 10 hours. 
Statistical estimation results are shown in Table~\ref{table:PMC_table}. The probability to the satisfaction of Property~\ref{sec:result:rq3:p1} increases along with the periods. Finally, the probability reaches 92\% when the period is 5000. According to the engineer's experience, this probabilistic increase is reasonable. The angle and the angular velocity will be smaller than a threshold after the period is greater than 5000.

Table~\ref{table:PMC_table} shows that the probability of the satisfaction of  property~\ref{sec:result:rq3:p2} is 0\% at all times. That means, the expected transition from \textit{mode m2} to other modes never occurs. This is obviously incorrect. By checking the specification, the practitioner found that this error was caused by a variable that was not correctly initialized. The system switched to \textit{mode m4} after finishing the tasks of \textit{mode m2} and finally got stuck in \textit{mode m4} due to this error. The experimental results convinced us that statistical estimation can detect more deep-rooted errors than human work. In our experiment, a total of 12 properties were built and verified. Requirement errors were detected by 3 properties while the rest 9 properties were satisfied.

\begin{table}[t]
    \centering 
    \caption{Results of the Statistical Estimation}
    \label{table:PMC_table}
    {
        \begin{tabular}{|c|c|c|c|}
            \hline
            Period&Property 1 &Property 2&Property 3\\ \hline
            {0} & {0\%}& {0\%}& {100\%}\\ \hline
            {500}& {0\%}& {0\%}& {100\%} \\ \hline
            {1000}& {0\%}& {0\%}& {100\%}\\ \hline
            {1500} & {4\%}& {0\%}& {100\%}\\ \hline
            {2000}& {11\%}& {0\%}  & {100\%} \\ \hline
            {2500}  & {28\%}& {0\%}& {97\%}\\ \hline
            {3000}& {36\%} & {0\%}& {83\%} \\ \hline
            {3500}& {48\%}& {0\%}& {64\%} \\ \hline
            {4000}& {59\%}& {0\%}& {46\%}\\ \hline
            {4500}& {73\%}& {0\%}& {35\%} \\ \hline
            {5000}& {92\%} & {0\%}& {27\%}\\ \hline
        \end{tabular}
     \vspace{-15pt}
    }
\end{table}

\begin{tcolorbox}\noindent\textbf{\emph{G4}}: How useful for the test-case generation under the MC/DC coverage to improve the testing? \end{tcolorbox}
We implemented the test-case generation under the MC/DC coverage in the aviation system. We generated 2,341 test cases in total from the formal specification. To validate the implementation w.r.t. the requirement, we ran our test cases on the source-code level. Summarily, 1,934 of the test cases could be directly used for testing, while the other 407 could not. The reasons are mainly that the requirements are: 1) incomplete, e.g., variables are missing; 2) imprecise, e.g., the fact of the minimum value being less than the maximum value is not an explicit statement; 3) inconsistent to the implementation, e.g., the requirement requires that the first period must perform differently with another one, but the implementation does not care about the first period.

In general, executing the test cases under the MC/DC coverage detects errors in both source codes and requirements. The automated generation of such test cases from requirements also accelerates the testing on the implementation stage. 

\begin{tcolorbox}\noindent\textbf{\emph{G5}}: How useful for the whole approach in practice, compared to the situation where it is not applied?  \end{tcolorbox}
To compare \approach and traditional technologies in the industry, we investigated some important subjective data from the 12 practitioners who were involved in this project. The investigation aims at evaluating the \textit{usability} of~\approach.
We designed a comprehensive questionnaire including 19 questions. Some questions were purely subjective, such as ``\textit{Do you feel that the diagrams really help you find errors}?'' , ``\textit{Are the time costs of simulation and statistical estimation acceptable? }'' and ``\textit{Is it difficult to write the LTL properties for statistical estimation?}''.  

On average, 8 fatal requirement errors have been detected for each system tested in our experiment. For all the nine projects of our partners, about 5-10 requirements errors could be discovered, depending on the projects. The time cost of modeling and V\&V has been reduced by more than 50\%, from 4-6 months to 2-3 months. On average, the time cost for a requirement engineer to finish the diagram-based review is about 1-2 weeks. If the specification is manually reviewed, it needs 6-8 weeks by each practitioner. 
The efficiency of test case generation has been improved since the test case generation is fully automated.The most controversial issue lies in statistical estimation. 10 of the 12 practitioners complained that the construction of LTL properties was difficult. The lack of the mathematical notations took them a lot of time to translate the domain properties to the LTL properties. This problem will be considered in our future work. To sum up, the statistics show that \approach is useful to most practitioners. 

\section{Experience and Lessons}\label{sec:experience_lesson}
In this section, we share our experience and lessons gained from the 8-year long projects with our partners in both the aerospace and aviation domains. 
We hope the community can benefit from our positive experience and get rid of the wasted efforts from our negative lessons.

\subsection{Experience}
One important experience is that researchers in formal methods must help engineers focus on \textbf{domain knowledge} rather than notation details. When our research project started eight years ago, we recommended the practitioner’s traditional formal methods with great confidence and ambitions. We believed that most of their problems with precise requirements and rigorous V\&V could be solved by formal methods. However, after a three-months practice, the domain practitioners gave up since they still could not figure out how to represent the domain-specific knowledge and features in a formal specification. Moreover, how to build the formal specification from scratch was not solved as well. To tackle this problem, we focused on developing a targeted formal engineering approach to the domains. The experimental results showed that the dedicated formal notations and the template-guided formal modeling process are more attractive to the practitioners, as they can get rid of complex formal notations and rules. The well-designed template can guide them to write the appropriate requirement specifications.

Another experience is the concept of "\textit{engineering}". The engineering processes and specific technologies are important for applying formal methods to the industry. Formal methods can be significantly helpful if they are properly implemented. Some formal verification techniques, such as theorem proving~\cite{davis1962proving}, may not be suitable for industrial practitioners since the training cost is too high. In fact, the lack of engineering approaches is the main reason that stops practitioners from applying formal methods to their projects. An ideal engineering approach should organize engineering processes coherently and systematically to tell the practitioners \textit{what to do} and \textit{how to do}. 
These experiences give us the hint that developing domain-specific engineering methodologies is more important than introducing general theories as well as research-oriented tools. 

Since 2012, \approach have been used in industrial projects to help detect requirement problems in critical systems. We verify the software in the requirement phase rather than design phase, and find many errors in the test-adequate requirement. However, We still can't get the useful properties to detect the deep-level errors, 
and our requirement language is limited for the specific domain. Furthermore fills requirement in our template is a labor-cost work. We will address these limitation in the future work with the help of the natural language processing technology and industry norms.

\subsection{Lessons}
 \noindent\textbf{Do not ask the engineers to learn complicated formal notations or proof knowledge.} We spent almost three months introducing formal notations and proofs to the engineers but failed. A more promising way may be to develop a dedicated approach rather than showing formal theories only.

\noindent\textbf{Do not make aggressive changes to the engineers' custom to modeling and V\&V}. The ways of requirement modeling and V\&V, for instance, writing specification with Microsoft Word, have been used by engineers for decades. Any new approach should not significantly change these traditional ways. A more rational idea is to integrate traditional ways in the new approach. That is, "\textit{reform}" is more appropriate than "\textit{revolution}". As a result, we design a Microsoft Word requirements template for formal specification construction.

\noindent\textbf{Never underestimate the importance of the tool support}. If the automation is insufficient, few engineers are willing to use the formal method. After a thorough investigation, we gradually developed our tool \tool to support the automation of \approach, which makes the engineers happier to use formal methods. This philosophy is similar to Karl Marx's "\textit{The weapon of criticism can certainly not replace the criticism of weapons}". Engineering problems need to be solved by powerful tools rather than only by theories.

 \section{Related Work}~\label{sec:rel_work}
More and more industry stakeholders want to improve software quality. \cite{planning2002economic} reports that at least 70\% of errors are introduced during the specification process and before implementation efforts. This indicates the requirement is the core stage of the software life-cycle. 

A good requirement-engineering process includes requirement elicitation, requirement modeling,  and analysis, requirement validation and negotiation~\cite{sommerville1997requirements}.
KAOS is a classic goal-oriented approach defining systems within agents, objects, operations to elicit requirements~\cite{kaos1993Dardenne}. The approach i*~\cite{yu1997towards} which focuses on modeling environment and features for the early-phase of requirement engineering is also used for requirement elicitation. However, our work focus on modeling and analysis requirement,  the start point of our work from the stated requirement from the industry. Leveson et al.~\cite{leveson2004STPA} proposed a systemic STPA method for accident analysis, hazard analysis, and accident prevention strategies. There is subsequent work on STPA, e.g.,~\cite{ishimatsu2010modeling,pereira2019stamp}, showing that STPA has a great improvement in the accident-oriented system. However, STPA is an accident-oriented modeling method, and company has to re-organize its development process from scratch to adopt STPA. Meanwhile, not every system is suitable for accident-oriented modeling. In fact, the industry engineers want an approach integrated into their existing development process.~\approach proposes the proper V\&V technologies meet this goal. Kafali et al.~\cite{kafali2016nane} propose a temporal reasoning framework NANE. NANE can help analysts identify misuse cases by formal reasoning about norm enactments, however, the process of norm extraction from the requirement is still a problem which is a general difficulty problem for modeling requirements.~\approach provides the language template to tackle this challenge.

In the aviation domain, Miller et al.~\cite{miller2003flight} propose a requirement specification written in the RSML-e language for the mode logic of the Flight Guidance System of a typical regional jet aircraft. However, the RSML-e language only focuses on verifying system modes without considering the verification of the computation task. Feiler et al.~\cite{feiler2006sae} proposed the Architecture Analysis \& Design Language(AADL) support early and repeated analyses of the embedded system. As an international standard (AS5506A~\cite{as55062004architecture}), AADL is wildly used in the aviation domain, like~\cite{delange2009validate,cofer2018formal}. In many projects, AADL is more suitable for system architecture modeling and validation than for requirement modeling and analysis. The reason is that system architecture defines "how to do " but requirement specifications focus on "what to do". Moreover, AADL is a semi-formal notation, which cannot support formal specification construction and analysis. Extant commercial tools, e.g., \emph{DOORS}~\cite{web2019Doors} and \emph{RTCASE}~\cite{web2019RTCASE} fall short in precise specification modeling and rigorous V\&V.

Several techniques of formal methods can be utilized to model requirements, e.g., the Z~\cite{jacky1997way}, B~\cite{abrial2005b}, and Event-B~\cite{abrial2010modeling} methods. These methods can model systems not only with physical environments but also with human users~\cite{abrial2018b}. To enrich the expressiveness, traditional state machines are improved for requirements modeling, such as the ASM~(Abstract State Machine) Method~\cite{borger2010abstract}. ~\emph{StateChart} is widely used in industrial software modeling and analysis~\cite{harel1987statecharts}. MatLab/SimuLink provides the~\emph{Stateflow} that is improved based on~\emph{StateChart}. The~\emph{SOFL} method that has been applied in domains like the services computing~\cite{miao2012formal} and the railway control systems~\cite{miao2016automated}, focuses on the requirement modeling and static analysis~\cite{liu1998sofl,liu2013formal,li2015integrating, huang2019prema}. Dietl et al.~\cite{dietl2012verification} automatically converts the program and property into a game that can be played by people with no knowledge of or training in computing. Dietl thinks that labor costs have heretofore made formal verification too costly to apply beyond small critical software components. The availability of inexpensive formal verification could change the economics of software V\&V. Morisio et al.~\cite{morisio2000extending} propose a few extensions to express variability, and they define precisely their semantics so that a tool can support them. 

\section{Conclusion}
\label{sec:conclusion}
\par In this paper, we present a formal engineering approach \approach to the formal requirements modeling and V\&V in the aerospace and aviation control domains. The specification construction is guided by a template that is developed based on domain knowledge and features. The requirements V\&V consists of the diagram-based review, requirement simulation, statistical estimation, and the MC/DC test-case generation techniques. We also have developed a supportive tool \tool for \approach provides users with a unified requirements specification construction and V\&V environment. So far, \approach has been applied to real aerospace and aviation control systems from our industrial partners, the AECC Commercial Aircraft Engine co. LTD and the China Academy of Space Technology. The experimental results demonstrate the feasibility and advantages of \approach. We have also reported the experience and the lessons when using the approach. In future work, we plan to apply the natural language processing techniques to automatically extract the formal specifications from user requirements. Improving our V\&V performance is also one of our research topics in the future.

\begin{acks}
    \par We thank the anonymous reviewers for their valuable
    feedback. Weikai Miao is supported by the NSFCs of China (No. 61872144 and No. 61872146). Geguang Pu is supported by NSFC Project No. 61632005 and NSFC Project. No. 61532019.
\end{acks}

\bibliographystyle{ACM-Reference-Format}
\bibliography{reference}

\end{document}